
\input manumac
\twelvepoint
\def\tr{{\rm tr}}
\def\M{{\cal M}}
\line{\hfil DAMTP/R-93/18}

\vskip 0.4truecm
\centerline{\bf AXIONIC DEFECT ANOMALIES AND THEIR CANCELLATION}
\vskip 1truecm
\centerline{\bf J.M. Izquierdo and P.K. Townsend}
\bigskip
\centerline{DAMTP, University of Cambridge}
\centerline{Cambridge, England}

\vskip 1cm

\AB
We present a simple derivation of the Callan-Harvey-Naculich effect, {\it
i.e.} the compensation of charge violation on axion strings due to gauge
anomalies by accretion of charge onto the string from the surrounding space. We
then show, in the case of axion fields without a potential, that an
alternative explanation is possible in which no reference to the surrounding
space is necessary because the anomalies are cancelled by a version of the
Green-Schwarz mechanism. We prove that such an alternative explanation is
always possible in the more general context of p-brane defects in
d-dimensional field theories, and hence that there always exists an
anomaly-free effective worldvolume action whenever the spacetime theory is
anomaly free. Our results have implications, which we discuss, for heterotic
and type II fivebranes.
\AE

\vfill\eject

\noindent {\bf 1. Introduction } \medskip
Let a
complex scalar field $S+iP=\rho e^{i\theta}$ be coupled to fermions $\psi$,
in a four-dimensional spacetime ${\cal M}$, through a term of the form $$
\bar\psi(S+\gamma_5 P)\psi \eqno (1.1)
$$
and let these fermions be coupled to a Yang-Mills (YM) one-form
potential $A$ in an anomaly-free representation. If the scalar field
potential is minimised when $\rho=m\ne0$, then at an energy scale much less
than $m$ the fermions can be integrated out to arrive at an effective action
for $\theta$ and $A$. Since $\theta$ is the Goldstone field for a global
$U(1)$ chiral invariance that is spontaneously broken when $\rho\ne0$, one
might expect it to appear in this effective action only through its derivative
$d\theta$. However, there is an anomalous triangle diagram that
produces an interaction term of the form
$$
S_{ax}={k \over 8\pi^2} \int_{{}_{\cal M}}\theta\, {\rm tr}(F\wedge F)\ ,
\eqno (1.2)
$$
where $F=dA +A^2$ is the Lie-algebra valued YM field-strength two-form, and
$k$ is an integer. Observe that this is consistent with periodicity in
$\theta$ because $S_{ax}[\theta +2\pi]$ differs from $S_{ax}[\theta]$ by
$2\pi$ times an integer. A field $\theta$ with a coupling of the form (1.2)
that is also periodically indentified is called an axion field. This field
may have string-like defects, called axion strings, around which $\theta$
changes by an integral multiple of $2\pi$.

It was shown by Callan and Harvey [1] that the Dirac equation for $\psi$ has
zero modes in the presence of an axion string. The coefficients
of these modes represent {\it chiral} fermions trapped on the string. Since
these fermions couple to the YM field they contribute to the YM current on
the string but, since they are chiral, this current is anomalous. At first
sight this is puzzling because the four-dimensional theory is anomaly free.
As Callan and Harvey pointed out, the resolution of this puzzle is that the
term (1.2) also contributes to the YM current in the neigbourhood of an axion
string (where $\theta$ is spatially varying) and, after taking into account
an important correction of Naculich [2], this current, which we shall refer to
as the CHN current, causes an accretion or depletion of charge on the string
from the surrounding space. This effect is found to precisely account for the
violation of charge conservation on the string due to the worldsheet
anomaly.

This resolution of the puzzle would seem to suggest that it would
be futile to seek an effective action for the axion string alone in a YM
background because any such action would fail to be invariant under
gauge-transformations of the background. This is indeed the case when the
axion field has a potential $V(\theta)$ (periodic in $\theta$ with period
$2\pi$) because the string is then the boundary of a domain wall and the
violation of charge on the string is due to currents in the wall induced by
the external YM field [3]. The CHN effect can thus be seen to
have its origin in effects previously described (in the abelian
case, at least) for axion domain walls [4,5,6]. Clearly, the
dynamics of the string cannot be separated from that of the wall of which it
is the boundary, so it would not be reasonable to expect there to exist a
gauge-invariant action for the string alone.

However, the case considered in [1] was that of an axion field {\it without} a
potential term, for which there is no domain wall. The situation in this
case is quite different because it is then possible to replace the axion field
by its dual, a two-form potential $B$ with an `anomalous' YM transformation.
The combined action for the spacetime fields and the string can now be split
into a spacetime part and a string part, each of which is {\it separately}
gauge-invariant. The string action found in this way is anomaly-free because
of the anomalous transformation of $B$. This is essentially a `sigma-model'
version (because the gauge-fields are pull-backs from spacetime) of the
Green-Schwarz (GS) anomaly cancellation mechanism (cf. [7]). There are
therefore two apparently very diferent ways to explain the cancellation of
anomalies of axion strings. According to the CHN mechanism the current
on the string really is anomalous but this is compensated by  a current in the
surrounding space. According to the alternative mechanism there is no need to
consider the surrounding space because the effective string action is
non-anomalous once its coupling to $B$ is included. This presents
us with another puzzle: how can both mechanisms be correct? We resolve this
puzzle by showing that there is an intrinsic ambiguity in the definition of the
electric or YM current which allows for a redistribution of charge without
changing the total charge.

There is a natural generalization of the axion field $\theta$ to a
(d-p-3)-form potential $b$ in a d-dimensional spacetime. The non-vanishing of
$db$ over some (d-p-2)-cycle in space signals the presence of a p-brane
defect, which will couple to the (p+1)-form dual potential $\tilde b$.
Again, the effective action of the defect may be chiral and hence potentially
anomalous; recent examples are the `elementary' fivebrane solutions of d=10
supergravity theories [8,9]. A question of considerable interest is whether it
is possible to find a gauge-invariant effective action for these objects in the
presence of spacetime gauge fields. This is not obviously guaranteed because
an anomaly of the effective action could always be explained as the result of
CHN-type currents in the surrounding space. We shall prove that whenever it is
possible to account for anomalies of the defect's effective action by the CHN
mechanism then it is also possible to cancel them by the sigma-model GS
mechanism. Hence, a gauge-invariant effective action for a p-brane defect
always exists provided that the d-dimensional field theory is anomaly free. We
conclude with a discussion of the implications of our results for the
heterotic and type II fivebranes.

\bigskip
\noindent {\bf 2. CHN mechanism in the dual formulation}
\medskip

We shall begin by presenting a simplified derivation of the results of CHN.
We first replace $\theta$ by its `dual', a two-form potential $B$. The
resulting
action for $A$ and $B$ is
$$
S_{\cal M} = \int\!\! d^4x\big\{ -{1\over4}\tr( F_{\mu\nu}F^{\mu\nu}) -
{1\over 12m^2}H_{\mu\nu\rho}H^{\mu\nu\rho}\big\}
\eqno (2.1)
$$
where
$H_{\mu\nu\rho}$ are the components of the field-strength three-form
$$
H= dB + {k\over 8\pi^2} \tr\big(AdA +{2\over3}A^3\big)\ .
\eqno (2.2)
$$
This is YM invariant provided that $B$ has the `anomalous' YM transformation
$$
\delta_\epsilon B = -{k\over 8\pi^2}\tr (Ad\epsilon)
\eqno (2.3)
$$
where $\epsilon$ is the Lie-algebra valued parameter of YM transformations.
The Chern-Simons (CS) term appearing in this field-strength has its origin
in the axion coupling (1.2) [10]. Note that the integral over spacetime of the
Bianchi identity for $H$ yields
$$
\int_{\cal M}\!\! dH \ =\ k\nu
\eqno (2.4)
$$
where $\nu$ is the integer instanton number of $A$.

Axion strings are incorporated in this dual approach by regarding them as
sources for $B$. Let $w$ be the worldsheet of an axion string and  $\phi :
w\to {\cal M}$ the map that specifies the embedding of $w$ in ${\cal M}$.
We add to the action the interaction term
$$
2\pi \int_w \!\! \phi^*(B)
\eqno (2.5)
$$
where $\phi^*(B)$ is the pullback of $B$ to $w$.

It might be thought
that the information contained in the identification $\theta\sim\theta
+2\pi$ has been lost by passing to the dual formulation in terms of $B$,
but this is not so. The significance of the periodic identification of
$\theta$ is firstly the possibility of axion strings, which we have just
taken into account, and secondly the integer quantization of $k$. This
quantization condition also arises in the dual approach as follows.
Since $B$ is defined only locally we rewrite (2.5) in terms of the
globally defined invariant field strength $H$ as
$$
2\pi \int_\Sigma
\!\! H\ - \  {k\over 4\pi} \int_\Sigma \!\! \tr\big(AdA +{2\over3}A^3\big)\ ,
\eqno (2.6)
$$
where $\Sigma$ is any three-manifold with $w$ as its boundary (we assume the
existence of $\Sigma$). But the second, CS, term in this expression is not YM
invariant for two reasons. Firstly because of the boundary, but this can be
dealt with by a boundary term to be introduced later. For now we can avoid this
problem by considering the special case for which $\partial\Sigma=0$. Then, as
is
well-known, the CS term is invariant under `small' gauge transformations but
not
`large' ones and this fact requires that $k$ be an integer [11].

It is convenient to take into account the chiral fermions on the axion string
by non-abelian bosonisation [12]. Specifically, we replace the fermions by a
`sigma-model' version (i.e. with gauge-fields induced from spacetime) of a
gauged WZW term for the YM group $G$ [13,14]. Let $L$ be the Lie-algebra
valued left-invariant one-forms on $G$, and let $h=\tr(L^3)$. Since $dh=0$ we
have, locally, that $h=db$ for two-form potential $b={1\over2}dx^\mu dx^\nu
b_{\mu\nu}$ on $G$. The worldsheet fermion action can now be replaced by
$$
-{k\over 4\pi} \int\!\! d^2\xi\, \varepsilon^{ij}\big[\tr(L_iA_j) +{1\over 6}
b_{ij}\big] \eqno (2.7)
$$
where $L_i$ and $b_{ij}$ are the components of the pullbacks to the worldsheet
of the forms $L$ and $b$ on $G$, respectively.

Now, let $y^m$ be coordinates for $G$, $g_{mn}$ the left-invariant
metric on $G$, and let $L^m\partial/\partial y^m$ be the Lie-algebra valued
left-invariant vector field dual to the one-form $L$. Taking all the above
into account, the effective string action can be taken to be [15,3]
$$
\eqalign{ S_{string} = \int\!\! d^2\xi\,
\bigg[ &- \sqrt{-\gamma}\gamma^{ij}\big(\partial_i x^\mu\partial_j x^\nu
g_{\mu\nu} +(D_iy)^m (D_jy)^n g_{mn}\big)\cr
&\qquad\qquad\qquad +
\pi \varepsilon^{ij}( B_{ij} -{k\over 4\pi^2} \tr(L_iA_j) -{k\over 24\pi^2}
b_{ij})\bigg]\cr}
\eqno (2.8)
$$
where $\gamma_{ij}$ is the worldsheet metric, and $(D_i y)^m$ are the
components of the pullback from ${\cal M}\times G$ of the YM covariant
derivative
$$
(Dy)^m = dy^m - \tr(A L^m)\ .
\eqno (2.9)
$$
The total action is
$$
S_{total} = S_{\cal M} + S_{string}
\eqno (2.10)
$$

We now have an adequate framework within which to provide a simple
derivation of the results of [1,2]. The essence of the point of view of
these references is that $B$ is to be regarded as a `matter' field
contributing to the YM current. Variation of the $H^2$ term in $S_{\cal M}$
in the {\it absence} of an axion string source yields the current
$$
J^\mu(CH)=-{k\over 8\pi^2m^2} H^{\mu\nu\rho}F_{\nu\rho}\ .
\eqno (2.11)
$$
Apart from the fact that we are using here the dual, two-form, formalism, the
current $J^\mu(CH)$ is that which was originally suggested [1] to account for
charge violation on the string due to the worldsheet anomaly. However, in
the presence of an axion string source there is an additional term [2] in the
current. Using the $B$ equation of motion
$$
\partial_\rho H^{\mu\nu\rho} = -m^2 J^{\mu\nu}\ ,
\eqno (2.12)
$$
where
$$
J^{\mu\nu}(x) \equiv 2{\delta S(string)\over \delta B_{\mu\nu}(x)} =2\pi
\int\!\!
d^2\xi\, \varepsilon^{ij}\partial_i X^\mu\partial_j X^\nu \,
\delta^{(4)}\big(x-X(\xi)\big) \ ,
\eqno (2.13)
$$
the current in the presence of an axion string source is found to be
$$
J^\mu(CHN)= J^\mu(CH) + J^\mu(N)
\eqno (2.14)
$$
where $J^\mu(N)$ is localized on the string. Any such localized current
$J^\mu$ can be written in terms of a worldsheet current $j^i$ as
$$
J^\mu(x) = \int\!\! d^2\xi\, j^i\, \partial_i X^\mu \,
\delta^{(4)}\big(x-X(\xi)\big) \ ,
\eqno (2.15)
$$
and its covariant divergence as
$$
D_\mu J^\mu (x) =\int\!\! d^2\xi\, D_i j^i\, \delta^{(4)}\big(x-X(\xi)\big) \
. \eqno (2.16)
$$
For the case in hand,
$$
j^i(N) = -{k\over 4\pi}\varepsilon^{ij}A_j\ .
\eqno (2.17)
$$

The current $J^\mu(CHN)$ is covariantly
conserved except on the string worldsheet. In fact, using (2.11) we find that
$$
D_\mu J^\mu(CHN) = {k\over 4\pi}\int\!\! d^2\xi\, \varepsilon^{ij}\partial_i
A_j \, \delta^{(4)}\big(x-X(\xi)\big)\ .
\eqno (2.18)
$$
Observe that both terms in (2.14) contribute to this result, and this is the
origin of the factor of two error in [1] that is corrected in [2].

We turn now to the current
$$
J^\mu(string)(x) \equiv {\delta S_{string}\over\delta A_\mu(x)}
\eqno (2.19)
$$
due to the massless modes on the string. This current is of course
localized on the string. All that need concern us about it here is that the
corresponding worldsheet current $j^i(string)$ is anomalous:
$$
D_i j^i(string) = -{k\over 4\pi}\varepsilon^{ij}\partial_i A_j\ .
\eqno (2.20)
$$
The right hand side of (2.20) is the `consistent' anomaly for a two-dimensional
YM theory. It was pointed out in [2] that since  $J^\mu(N)$ is localized on
the string one can consider the corresponding worldsheet current $j^i(N)$ as
part of a `total' worldsheet current
$$
j^i(total)= j^i(string) +j^i(N)\ .
\eqno (2.21)
$$
This current is still anomalous,
$$
D_ij^i(total) =-{k\over 4\pi}\varepsilon^{ij}F_{ij}\ ,
\eqno (2.22)
$$
but the anomaly is now {\it covariant}. With either definition of the
worldsheet current, the {\it total} spacetime current
$$
J^\mu = J^\mu(CHN) + J^\mu(string)
\eqno (2.23)
$$
is covariantly conserved. This completes our derivation of the results of
[1,2].

\bigskip
\noindent {\bf 3. Anomaly cancellation by the GS mechanism}
\medskip

We shall now present
another resolution of the puzzle of axion string anomalies that is, on the
face of it, entirely different. The essence of the new point of view, which is
implicit in [3] but was not fully appreciated there, is that $B$ is not to be
treated as a `matter' field for $A$ but as a gauge field on the same footing
as $A$.
We have seen that the spacetime action (2.1) is invariant under the gauge
transformations
$$
\eqalign{ \delta A
&= d\epsilon +[A,\epsilon]\cr
\delta B  &= -{k\over 8\pi^2}\tr(Ad\epsilon) + d\Lambda}
\eqno (3.1)
$$
where $\Lambda$ is a one-form parameter of the abelian gauge transformation
of $B$. The string action (2.8) is also invariant (up to surface terms [16])
under these transformations if they are supplemented by the additional
transformation $$
\delta_\epsilon y^m = \tr(L^m\epsilon)\ .
\eqno (3.2)
$$

In effect, we have just described a bosonized version of the standard
`sigma-model anomaly' cancellation due to an `anomalous' transformation of $B$
[7]. However, since there is an apparent contradiction with the CHN point of
view in which gauge invariance requires consideration of the surrounding
spacetime, a more detailed explanation of the consistency of the new point of
view may be useful. Following the suggestion in [3] for the abelian case we
now take the YM current of the string to be
$$
\tilde J^\mu(string) (x) = {\delta S_{string}\over\delta A_\mu(x)} +{k\over
4\pi^2} A_\nu(x){\delta S_{string}\over\delta B_{\mu\nu}(x)}\ ,
\eqno (3.3)
$$
for which the associated worldsheet current is
$$
\eqalign{
\tilde j^i(string) &= 2\sqrt{-\gamma}\gamma^{ij}D_jy
+{k\over4\pi}\varepsilon^{ij}\partial_j y + {k\over4\pi}
\varepsilon^{ij}A_j\cr
&= j^i(string) + {k\over 4\pi}\varepsilon^{ij}A_j\cr
&= j^i(total) +{k\over 2\pi}\varepsilon^{ij}A_j\ .}
\eqno (3.4)
$$
It is instructive to examine the difference between the two currents $j(total)$
and $\tilde j(string)$. Consider first the abelian case. In this case the
current $j(total)$ is anomalous but {\it invariant} while the new current
$\tilde j(string)$ is conserved but appears to be gauge non-invariant. However,
the algebra of gauge transformations of $A$ and $B$ is non-abelian even for a
$U(1)$ gauge group because a commutator of two $U(1)$-transformations on $B$
produces an antisymmetric tensor gauge transformation, i.e.
$$
[\delta_{\epsilon_1},\delta_{\epsilon_2}] =
\delta_{\Lambda= 2k\epsilon_2
{\buildrel \leftrightarrow\over {\partial_\mu}}\ \epsilon_1}\ .
\eqno (3.5)
$$
It is known from previous work [15,16] that the generators of $U(1)$ gauge
transformations on string wavefunctionals take the form
$$
\oint \!\! d\sigma\, \epsilon\big(x(\sigma)\big) D(\sigma)
\eqno (3.6)
$$
where $\sigma$ is the (closed) string coordinate and the hermitian operators
$iD(\sigma)$ obey a $U(1)$ Ka{\v c}-Moody algebra {\it with central
extension}. In fact, $$
D(\sigma) = {\delta\over \delta y(\sigma)} - i{k\over4\pi} y'
\eqno (3.7)
$$
where the prime indicates differentiation with respect to $\sigma$. This may
now be compared with the charge density $\tilde j^0(string)$, which can be
written in terms of the momentum $\Pi_y$ conjugate to $y(\sigma)$ as
$$
\tilde j^0(string) = -\Pi_y + {k\over4\pi} y'\ .
\eqno (3.8)
$$
Hence, upon canonical quantization, $\Pi_y(\sigma) \to -i\delta/\delta
y(\sigma)$, we have that $\tilde j^0(string)\to iD(\sigma)$, and we conclude
that $\tilde j^i(string)$ has the {\it covariance} properties required of
currents associated with the gauge fields $A$ and $B$ with gauge algebra
(3.5). Similar arguments confirm also in the non-abelian case that $\tilde j$
is the appropriate current despite its apparent non-covariance.

We are now faced with a new puzzle: given that both points of view on
anomalies of axion defects are mathematically self-consistent, how can the
apparently different physics of these two points of view be reconciled?
To answer this it will prove useful to consider again the abelian gauge group
$U(1)$. In this case the spacetime current $J^\mu$ of (2.23) is conserved
but, unlike $\tilde J^\mu (string)$, it is not localized on the string.
However, $J^\mu$ suffers from an intrinsic ambiguity because the current
$$
\eqalign{ J'{}^\mu &= J^\mu - {\alpha\over m^2}\partial_\nu \big(
H^{\mu\nu\rho} A_\rho\big)\cr
&= \big[J^\mu -{1\over 2m^2}\alpha H^{\mu\nu\rho}
F_{\nu\rho}\big] -{\alpha\over m^2} \big(\partial_\nu H^{\mu\nu\rho}\big)
A_\rho}
\eqno (3.9)
$$
is also conserved for any value of the dimensionless parameter $\alpha$ and
yields the same total charge provided only that the gauge fields tend to zero
sufficiently fast as spatial infinity is approached; a condition that is
satisfied by the fields due to a closed string confined to any sphere of
finite radius. Note that the additional term in the current is identically
conserved, while the conservation of $J^\mu$ depends on the use of the $B$
field equation (because in this approach it is regarded as part of the
`matter'). We may therefore use this field equation to rewrite $J'{}^\mu$ as
$$
J'{}^\mu = \big[J^\mu -{1\over 2m^2}\alpha H^{\mu\nu\rho} F_{\nu\rho}\big] -
{2\alpha\over m^2} J^{\mu\nu}A_\rho\ .
\eqno (3.10)
$$
The last term on the right hand side of this equation is localized on the
string. The sum of the first two terms is not in general localized on the
string but for the choice of $\alpha= -{k\over 4\pi^2}$ it {\it is}. In
fact, for this choice of $\alpha$ we find that
$$
J'{}^\mu =\tilde J^\mu \ .
\eqno (3.11)
$$
We conclude that the difference between the currents in the two approaches to
anomaly cancellation is an identically conserved current which does not
contribute to the total charge but which does redistribute it. The
requirement that the charge be localized on the string fixes uniquely this
ambiguity. A similar analysis can be made for
non-abelian currents provided that one adds to both  $J'{}^\mu$ and $\tilde
J^\mu$ the contribution $[A\nu , F^{\mu\nu}]$ of the YM field (so as to be
able to compare conserved, rather than covariantly conserved, currents).

\bigskip
\noindent {\bf 4. Anomaly cancellation for p-branes in d-dimensions}
\medskip

As we have shown, both the CHN mechanism and the sigma-model version of the
GS mechanism are equally valid ways of understanding the cancellation
of anomalies of axion strings for axion fields without a potential. As
emphasised
in [3], many of the recently found p-brane solutions of ten and
eleven-dimensional supergravity theories can be regarded as natural
generalisations of axion strings. Like axion strings the effective worldvolume
field theory of these defects is in many cases chiral and hence potentially
anomalous, although the spacetime field theory is anomaly free. Examples are
the fivebrane solution of the low energy ten-dimensional supergravity theory
for the heterotic string, and the fivebrane solutions of d=11 and type IIA
supergravity. In all of these cases anomalies
of the worldvolume field theory must be compensated by the CHN mechanism, but
we would like to know whether they can also be cancelled by the sigma-model
GS mechanism, because only in this case could one regard the worldvolume field
theory as a candidate for a fundamental p-brane theory. We now show that this
is indeed always possible, i.e. that whenever worldvolume anomalies are
compensated by the CHN mechanism then they can also be cancelled
by the anomalous variation, implied by the axion potential coupling,
of the (p+1)-form dual potential.

Let $b$ be a (d-p-3)-form potential (the analogue of $\theta$ in
the d=4, p=1 case) with the $({\rm d-p-2})$-form field-strength $h$. The
presence of a p-brane defect is signalled by the non-vanishing of the integral
of $h$ over some (d-p-2)-sphere in space. For example, this will happen if
$$
dh=\star J\ ,
\eqno (4.1)
$$
where $J$ is the (p+1)-form with components
$$
J^{\mu_1\dots\mu_{p+1}}(x)
= vol(S^{d-p-2})\int_w\phi^*\big(dX^{\mu_1}\wedge \dots\wedge
dX^{\mu_{p+1}}\big) \delta^d\big(x-X(\xi)\big)\ ,
\eqno (4.2)
$$
and $vol(S^n)=2\pi^{n/2}/\Gamma(n/2)$ is the volume of the unit n-sphere. The
symbol $\star$ indicates the Hodge dual in d-dimensional Minkowski spacetime. A
p-brane defect for which (4.1) holds will be called `ideal' because it is an
idealization of an actual defect in which the width of the core is taken to
zero. We defer to the next section a discussion of non-ideal defects.

We shall suppose that the effective spacetime action contains the interaction
term
$$
S_{ax}= -\int_\M \!\omega_{p+2}^{0} \wedge h \ .
\eqno (4.3)
$$
Here, $\omega_{p+2}^{0}$ is a Chern-Simons (p+2)-form such that
$$
d\omega_{p+2}^0 = X_{p+3}(F)
\eqno (4.4)
$$
where the (p+3)-form $X_{p+3}(F)$ is an invariant polynomial in $F$. The
existence of the interaction (4.3) is essentially what is meant by the
statement that $b$ is `axionic'. At this point we see that p must be odd.

In the discussion so far $F$ was a YM field-strength two-form but it can be
considered to include the curvature two-form of the d-dimensional spacetime,
i.e. {\it the results to follow apply equally to gravitational and mixed
gravitational/YM anomalies}. They may also be easily generalized to
the case in which $F$ is a form of degree higher than two.

To determine the YM variation of $S_{ax}$ we note first that
$$
\delta_\epsilon \omega_{p+2}^0 = d\omega_{p+1}^1(A,\epsilon)
\eqno (4.5)
$$
for (p+1)-form $\omega_{p+1}^1$. Hence
$$
\delta_\epsilon S_{ax} =-\int_\M \!\! d\omega_{p+1}^1\wedge h\ .
\eqno (4.6)
$$
Discarding a surface term at infinity we can rewrite this as
$$
\delta_\epsilon S_{ax} = \int_\M \!\! \omega_{p+1}^1\wedge dh\ .
\eqno (4.7)
$$
In the absence of a defect, $dh=0$ and $S_{ax}$ is therefore invariant. In
the presence of a defect we instead find that
$$
\delta_\epsilon S_{ax} = vol(S^{d-p-2}) \int_w\!\! \phi^*(\omega_{p+1}^1)
\eqno (4.8)
$$
Note that an alternative way to derive this result would be to insist on
$dh=0$ but to exclude from $\M$ the (p+1)-dimensional worldvolume of the
defect, $w$; this would produce an `inner' boundary and (4.8) would then
arise from the boundary terms neglected above in the integration by parts
that takes us from (4.6) to (4.7).

Since the original field theory is assumed to be anomaly free, the non-zero
variation of $S_{ax}$ must be cancelled by an equal but opposite variation of
the defect's effective action $S_{def}$. If, for example, the interaction term
$S_{ax}$ arose as a one-loop quantum contribution of fermion fields to the
spacetime effective action then we would expect the anomalous variation of
$S_{def}$ to be the result of chiral fermions trapped on the defect. Thus
$$
\delta_\epsilon S_{def} = -vol(S^{d-p-2})\int_w\!\!\phi^*(\omega_{p+1}^1)\ .
\eqno (4.9)
$$
Equivalently, (4.9) can be viewed as an anomaly of the effective worldvolume
action $S_{def}$ that is cancelled by the anomalous variation (4.8) of
$S_{ax}$. This is the standard CHN mechanism.

We shall now pass to the dual formulation in which $b$ is replaced by the
(p+1)-form $\tilde b$. We begin from the action
$$
S_h = \int_\M \big[  -{1\over 2} h\wedge \star h - \omega_{p+2}^0\wedge h
-\big(d\tilde b\wedge h +\tilde b\wedge \star J\big)\big]\ .
\eqno (4.10)
$$
In this form of the action $h$ is an {\it independent} (d-p-2)-form, and
$\tilde b$ is a Lagrange multiplier imposing the constraint (4.1). In the
absence of a defect this constraint is solved by setting $h=db$ and the
action reduces to the usual one (including the axion interaction term
$S_{ax}$). If $h$ is now eliminated from (4.10) by means of its field
equation,
$$
(-1)^d\star h =d\tilde b + \omega_{p+2}^0(A) \equiv \tilde h \ ,
\eqno (4.11)
$$
we find that
$$
S_h\to S_{\tilde h}= -{1\over 2}\int_\M  \tilde h\wedge \star\tilde h\
- vol(S^{d-p-2})\int_w\! \phi^*(\tilde b)\ .
\eqno (4.12)
$$
Using (4.5) one sees that $\tilde h$ will be YM invariant if
$$
\delta_\epsilon \tilde b = -\omega_{p+1}^1(A,\epsilon)\ .
\eqno (4.13)
$$
For this choice we find that
$$
\delta_\epsilon S_{\tilde h} =  vol(S^{d-p-2})\int_w\! \phi^*(\omega_{p+1}^1)\
, \eqno (4.14)
$$
which cancels (4.9). This is the CHN mechanism in the dual formulation.

Observe now that the anomalous variation of the `spacetime' part of the
action arises, in the dual formulation, entirely from the last term
in (4.12). But this term is {\it localized on the defect}. We may
therefore move it from the `spacetime' part of the action into the
defect's effective worldvolume action, {\it without changing the total
action}. When this is done the new `spacetime' action is obviously YM
invariant, as is the new worldvolume action,
$$
S_{def}' = S_{def} -vol(S^{d-p-2})\int_w\! \phi^*(\tilde b)\ .
\eqno (4.15)
$$
The worldvolume anomaly has now been cancelled
{\it without appeal to currents in the surrounding space}, in the sense that
the defect's effective action is both localized on the defect and
YM-invariant.

\bigskip
\noindent
{\bf 5. Anomaly cancellation for non-singular p-branes}
\medskip

There are three ways known to us in which an actual solution
representing a p-brane defect can have a non-singular core.
These are:
\medskip
(i) The form $b$ is singular in the action as given but this action is an
effective one valid only outside the core. The core width is determined by the
physics of a non-singular solution of the underlying field theory. This
is the case for the axion string example discussed previously.
\medskip
(ii) The field strength $h$ involves a CS term such that $dh= X(F)$ for an
invariant polynomial $X$ in a field-strength two-form $F$, and $\int_B
X(F)=\nu$ where $B$ is the (d-p-1)~-dimensional space transverse to the
defect's worldvolume, and $\nu$ is a non-zero topological index.
The core width is determined by the dimensions of the lowest energy
solution for given $\nu$. An example is the fivebrane
solution with YM instanton core of the field theory limit of the heterotic
string [17], (the heteotic fivebrane) or of d=10 supergravity/YM theory
[18].
\medskip
(iii) The form $h$ is closed but not exact. This can happen
only if the (d-p-2)-sphere surrounding the defect is a non-trivial
(d-p-2)-cycle in spacetime. An example is the self-dual threebrane [19,20] of
d=10 supergravity, which interpolates between Minkowski spacetime at infinity
and the product $(adS)_5\times S^5$ down an infinite wormhole throat [21].
\medskip

In all cases the actual defect can be effectively described
at wavelengths long compared to the core width by an `ideal' p-brane of zero
core width, so the analysis of the previous section should apply in all three
cases. However, it is instructive to consider in more detail how this comes
about. Here we shall consider case (ii), for which
$$
dh = X_{d-p-1}(F)\ .
\eqno (5.1)
$$
That is,
$$
h= db +\omega^0_{d-p-2}\ .
\eqno (5.2)
$$
where $d\omega^0_{d-p-2}= X_{d-p-1}(F)$. The action for $h$ will be taken to be
$$
S_h =-{1\over2}\int_\M \! h\wedge \star h \ -\ \int_\M \!
\omega^0_{p+2}\wedge h\ ,
\eqno (5.3)
$$
where $\omega^0_{p+2}$ is such that
$$
d\omega^0_{p+2}= X_{p+3}\ .
\eqno (5.4)
$$
We therefore require $p$ to be odd and $d$ to be even. We exclude from
consideration the case for which $h$ is a self-dual (d/2)-form since the
action (5.3) is then inappropriate. The
kinetic term of (5.3) is gauge-invariant provided that
$$
\delta_\epsilon b= -\omega^1_{d-p-3}\ ,
\eqno (5.5)
$$
where $\omega^1_{d-p-3}$ is defined as in (4.5). The complete action is not
gauge-invariant, however, because
$$
\delta S_h =\int_\M \! \omega^1_{p+1} \wedge X_{d-p-1}\ .
\eqno (5.6)
$$
For the complete theory to be anomaly free there has to be something that
compensates this anomaly. This will happen, for example, if there is an
anomalous fermionic part of the action that contributes a term equal to (5.6)
but of the opposite sign (as for the heterotic fivebrane).

We now suppose that there exists a static infinite p-brane field
configuration. That is, one for which the fields depend
only on the coordinates of the (d-p-1)-dimensional space $B$ orthogonal
to the (p+1)-dimensional worldvolume of the p-brane, and such that the
energy density is concentrated in a `core' region near the origin of $B$. We
further suppose that the gauge field configuration $A=\bar A$ is such that
$\bar A$ takes values in a  subalgebra ${\cal H}$ of the gauge algebra ${\cal
G}$, and that
$$
\int_{B}\!X_{d-p-1}(\bar A) =\nu \ne 0 \eqno (5.7)
$$
where $\nu$ is a non-zero topological index. The decomposition of the
adjoint represntation of ${\cal G}$ into representations of ${\cal H}$ will
generally contain singlets of ${\cal H}$, which span the adjoint representation
of another subalgebra ${\cal H}'$ of ${\cal G}$. For example, if ${\cal G}$
is $so(32)$ and ${\cal H}=so(3)$ then ${\cal H}'=so(29)$. Now let $A= \bar A
+a$, where $a$ is approximately constant across
the core and takes values in ${\cal H}'$. Then (5.6) yields
$$
\delta S_h
\approx \int_w\!\omega^1_{p+1}(a)\int_{B} \! X_{d-p-1}(\bar A)  = \nu
\int_w\!\omega^1_{p+1}(a) \ .
\eqno (5.8)
$$
The approximation becomes exact in the limit of zero core width. Thus, at
least for the subalgebra ${\cal H}'$ of ${\cal G}$, there is a contribution to
the anomaly that is localized on the worldvolume. Since the original theory was
assumed to be anomaly free as a result of including the fermions this has to
be cancelled by an equal but opposite contribution from chiral fermions
trapped on the object. In principle, it could be verified that these chiral
fermions exist (as a result of an index theorem) and that they contribute to
the anomaly in the required way. The above restriction on $a$ to the
subalgebra ${\cal H}'$ (which we do not see how to
relax\footnote{${}^*$}{although, presumably, the restriction to a
subalgebra does not arise for non-singular p-branes of type
(iii).}) suggests that these fermions should couple naturally
(locally?) only to gauge fields in this subalgebra. This is exactly what was
found in [18] for the  the heterotic fivebrane with YM gauge group $SO(32)$;
in this case there are 29 worldvolume fermions (as in the $E_8\times E_8$ case
[17]) which couple to the YM fields in the $so(29)$ subalgebra of $so(32)$. It
was verified in [18] that these 29 fermions do contribute to the sigma model
anomaly in the required way.

Our purpose here is to show that anomaly freedom of the spacetime theory
implies anomaly freedom of the effective worldvolume theory of the
defect\footnote{${}^{**}$}{The difference with the previous section, apart
from the fact that the defect is non-ideal, is that here one cannot
simultaneously have gauge-invariance and locality of the classical spacetime
action, although the discussion of the previous section can be generalized to
allow for this possibility too.}. To this end, we consider the dual
formulation. The dual action can be deduced as before, but now the Lagrange
multiplier term is  $\int_\M  (dh -X_{d-p-2})\wedge \tilde b$, which leads to
the result
$$
S_{\tilde h} = -{1\over2}\int_\M\! \tilde h\wedge \star\tilde h \ -\
\int_\M\! X_{d-p-1}\wedge \tilde b\ ,
\eqno (5.9)
$$
where $\tilde b$ has the anomalous transformation law
$$
\delta_\epsilon \tilde b = -\omega^1_{p+1}\ .
\eqno (5.10)
$$
The action $S_{\tilde h}$ has the same anomaly as $S_h$, so the two
formulations are equivalent.
Now suppose again that we have a p-brane defect with $A=\bar A$, and assume
that $\tilde b$ is approximately constant across the p-brane core. Then the
second term in (5.9) is approximately
$$
-\nu \int_w\!\tilde b\ .
\eqno (5.11)
$$
 From (5.10) we see that the variation of this term is exactly (5.8), as
before, but now the anomalous term in the action, i.e. (5.11), {\it and not
just its variation}, is localized on the worldvolume. Thus, this term can be
included in the p-brane's effective action which is then gauge-invariant
after taking into account the anomaly due to worldvolume fermions.

Again we see that if
the worldvolume anomalies due to chiral fermions are compensated by the CHN
effect (which is necessarily true for an anomaly free theory) then they can
also be cancelled by the sigma-model version of the GS mechanism.

\bigskip
\noindent
{\bf 6. Comments and Applications}
\medskip
In any anomaly-free field theory there is a correlation between `topological'
terms, such as $\varepsilon \theta FF$, in the the Lagrangian and anomalies
of axion defects. The correlation is required for the cancellation of these
anomalies. In the CHN approach the topological term contributes to the CHN
current, which accounts for charge violation on the defect. In the
alternative approach advocated here the topological terms imply an anomalous
gauge-transformation of the field dual to the axion, which allows the
construction of a modified definition of the gauge current on the defect and an
{\it invariant} effective action for it.

Given this correlation, a new puzzle arises on consideration of Lagrangians
with topological terms that did {\it not} arise from integration over
fermions. For example, one could simply add the $\varepsilon \theta FF$
term to the purely bosonic action for $\theta$ and $A_\mu$. This term would
then contribute to the CHN current in the presence of an axion string but now
there are no fermions and hence no chiral fermions trapped on the string.
The resolution in this case is simply that the axion string in this model is
singular. Of course, the spirit of the effective action approach is to replace
non-singular solutions with singular ones (since the difference is
unimportant at the macroscopic level) but not every singular solution
can be regarded as the idealization of a non-singular one, and only
those which can be are physically acceptable. Any more complete
model (e.g. in which $\theta$ appears as the phase of the complex field $\rho
e^{i\theta}$) for which the axion string is non-singular must yield an
anomalous effective action for the defect (adopting here the CHN point of
view) whenever the $\varepsilon \theta FF$ term appears in the effective
spacetime action.

The puzzle is more acute in supergravity theories for which topological terms
arise naturally in the classical action. As an example, consider the black
string solution [19] of the 6-form formulation of d=10 supergravity. Viewed as
a solution of d=10 supergravity/YM theory this string is axionic because of
the coupling $\int b_6\wedge {\rm tr}(F\wedge F)$ required by supersymmetry.
This term will cause an accretion of charge onto the string (for an
appropriate YM field configuration). This example might be excluded on the
grounds that it has fermions and is anomalous. We could surmount this
criticism by considering the low energy limit of the heterotic string in which
case we would have to include additional terms, specialize to the gauge
group $SO(32)$ or $E_8\times E_8$, and then apply the results of section 5,
but a much simpler anomaly-free example of relevance to our current discussion
is obtained by the simple expedient of throwing out the fermions. Thus, here is
an example of an anomaly free field theory with a string solution which must
be anomalous (from the CHN point of view) but with {\it no chiral fermions on
the string}. In this case one might question whether an effective description
of the string in terms of a worldsheet action is appropriate because, for
example, a black string has a non-zero temperature due to Hawking radiation
and cannot really be regarded as an `extended soliton'. Taking the zero
temperature limit does not help because the resulting `extreme' string
solution [22] is singular [19]. It might be thought that a better example
would be the extreme fivebrane solution of the bosonic sector of d=10
supergravity/YM theory [8,9], because this is non-singular; it
interpolates between d=10 Minkowski spacetime and a particular $S^3$
compactification with a linear dilaton [21]. Here however, the
axionic field $b$ is the two-form and its topological term  $\int b\wedge
X_8$ is absent from the d=10 supergravity/YM action unless one considers
the extension to the field theory limit of the heterotic string, but the
fermions are then required for spacetime anomaly freedom and the possibility
of chiral worldvolume fermions returns. In this case it is clear what
actually happens to any conserved charges accreting on the fivebrane; they go
down the infinite wormhole throat. The difficulty is in finding the correct
effective description of this process. This may involve the bosonic p-brane
action of [23] if worldvolume fermions are not required.

The type IIA d=10 fivebrane is also of interest because it provides a
further example of a defect in a non-anomalous, because non-chiral, field
theory for which the effective worldvolume field theory {\it is} chiral (the
physical field content is that of the d=6 antisymmetric tensor supermultiplet
[8]) and hence potentially anomalous. In fact, this remains true if the d=10
IIA supergravity is replaced by its purely bosonic sector; we must then
replace the effective worldvolume theory by its purely bosonic sector, i.e.
five scalars and one second rank antisymmetric tensor potential with {\it
self-dual} third-rank field strength. In principle, the antisymmetric tensor
could produce a `sigma-model' Lorentz anomaly because of the self-duality of
its field-strength. In fact it cannot do so because there would be no way that
this anomaly could cancel (recall that there are no fermions, by hypothesis,
and that none of the antisymmetric tensors of IIA supergravity has an
anomalous Lorentz transformation). Prior to the results reported here one
might have considered an inability to cancel worldvolume anomalies as simply
an indication that charge (or energy and momentum in this case) must accrete
onto the object from the surrounding space via the CHN effect. Now we know
that even if this explanation were possible (which it is not, because there are
no `topological' terms in the action that involve the curvature tensor) an
invariant effective action would still exist. Recall that in the above
discussion we discarded the fermions of the IIA supergravity theory. Let us
now reinstate them. Since the fivebrane action of the purely bosonic theory
is anomaly free, for the reason just explained, and since the action with
fermions is also anomaly free, the worldsheet fermions cannot produce  a
`sigma-model' Lorentz anomaly either, despite their worldvolume chirality.
This is not as surprising as it may sound because worldvolume fermions do not
couple universally to the {\it spacetime} Lorentz connection; In fact, they
may have no coupling to it.

The above remarks also have implications for `heterotic' fivebranes. In a
recent study of sigma-model anomalies for heterotic fivebranes [24] it was
argued that the standard formula for gravitational anomalies in six dimensions
(derived from the coupling to a six-dimensional metric [25]) could be used to
deduce the Lorentz (and mixed YM/Lorentz) sigma-model anomalies of the
fivebrane. This hypothesis contradicts the above conclusion concerning the
absence of these anomalies for the IIA fivebrane because both the
antisymmetric tensor and the fermions would contribute to the Lorentz anomaly
according to this hypothesis. Neither is it true that the two contributions
(computed in this way) cancel. Fortunately, there is no reason to believe the
hypothesis to be true. To see why, we review the arguments of [24] in favour
of it. Consider first the heterotic string. In this case the sigma model
anomalies have been computed [7], with a result that is summarized by the
anomaly 4-form
$$
X_4= -{1\over 16\pi^2}[\tr F^2 -\tr R^2]\ ,
\eqno (6.1)
$$
where the trace is to be taken in the vector representation.
This is to be compared with the result for standard (i.e. {\it
not} of `sigma-model' type) two-dimensional YM and Lorentz anomalies
for chiral and anti-chiral Majorana fermions. Let $r$
be the difference in dimensionalities of the YM representations of the chiral
and anti-chiral fermions. The anomalies may then be summarized by the anomaly
four-form
$$
I_4 =-{1\over 16\pi^2}[\tr F^2 -{r\over24}\tr R^2] \ .
\eqno (6.2)
$$
In [24] it is pointed out that if $F$ and $R$ in (6.2) are interpreted to
mean the pull-backs to the worldsheet of the {\it spacetime} YM field-strength
and curvature two-forms then (6.2) agrees with (6.1) for the heterotic
string because there are 32 heterotic femions of one chirality and 8 NSR
fermions of the other chirality, so $r=32-8=24$. This is undoubtedly a
curious fact but it bears no relation to the actual calculation, which shows
[7] that the {\it entire} contribution to the Lorentz anomaly comes from the
NSR fermions. The heterotic fermions make no contribution because they don't
couple to the spacetime Lorentz connection. This illustrates the
point, which is evident from comparison of (6.1) and (6.2), that although the
YM sigma-model anomaly has the same form as the `standard' YM anomaly this is
not true of the Lorentz anomalies. This is to be expected because whereas
worldsheet fermions couple universally to the worldsheet spin-connection they
may have quite different couplings to the spacetime connection.

For the heterotic fivebrane, the GS spacetime anomaly cancellation implies that
the six-form potential of d=10 supergravity/YM has an anomalous YM and
Lorentz transformation determined by cohomological descent from the
8-form
$$
X_8= {1\over 384\pi^4}[\tr F^4 -{1\over8}\tr F^2\tr R^2 +{1\over 32}(\tr
R^2)^2 + {1\over 8}\tr R^4]\ .
\eqno (6.3)
$$
Our results imply that this must be the anomaly 8-form
summarising the sigma-model anomalies of the six-dimensional effective field
theory of the fivebrane, although we cannot yet perform the calculation to
verify this because we do not yet have the complete spacetime
Lorentz-invariant worldvolume action. Nevertheless, in the spirit of ref.
[24], let us consider whether the results of such a calculation might be
deduced from the anomaly 8-form for the `standard' six-dimensional YM and
Lorentz anomalies. This 8-form is
$$
I_8= {1\over 384\pi^4}[\tr F^4 -{1\over4}\tr F^2\tr R^2
+{r\over 192}(\tr R^2)^2 + {r\over 240}\tr R^4] \ .
\eqno (6.4)
$$
There is agreement with (6.3) for the pure YM anomaly, again interpreting the
YM  two-form $F$ of (6.4) to mean the pull-back of the spacetime YM two-form.
If $r=30$ there is also agreement for the $\tr R^4$ part of the Lorentz
anomaly. It is argued in [24] that $r=32-2=30$ with the 32 coming from the 32
`heterotic' fermions of one chirality (as for the string) and the -2 coming
from the fermion partners to the four bosonic Goldstone fields arising from the
breaking of translation invariance in the four directions transverse to the
six-dimensional worldvolume. However, since there are no worldvolume gauge
fields in the effective action, all fermions must belong to six-dimensional
hypermultiplets and  must therefore all have the {\it same} chirality. Taking
into account a factor of 2 error, a correct calculation
based on the assumptions made in [24] would lead to the conclusion that
$r=32+1=33$. In fact, it may well be that $r=30$ is correct after all because
for the fivebrane solution with a YM instanton core it is known that the
number of heterotic fermions is actually 29 rather than 32 [17,18], so that
$r=29+1=30$. In this case a calculation of the YM sigma-model anomaly for an
$SO(29)$ subgroup of $SO(32)$ can be performed [18], and agreement is
found with (6.3). In contrast, the agreement between (6.4)
and (6.3) for the $\tr R^4$ term when $r=30$ is again a curious fact of no
obvious significance. Note that the gauge-fixed worldvolume action, for which
we at least know the field content, has no six-dimensional metric and is not
reparametrization invariant. Presumably, it can be made reparametrization
invariant by the inclusion of additional unphysical fields but the
contribution of these fields to (6.4) must then also be taken into account. We
expect that when this is done the total worldsheet gravitational anomaly will
be seen to vanish because any other result would be fatal to hopes of finding
a fundamental fivebrane theory dual to the heterotic string.

It should now be clear that there is no good reason for believing
that the formula $I_8$ can be used to calculate the gravitational and
mixed {\it sigma-model} anomalies of the heterotic fivebrane.
We are forced to conclude that, except for the pure YM anomaly, it is not yet
possible to confirm or refute by a worldvolume sigma-model anomaly calculation
our claim that the anomaly 8-form of the heteotic fivebrane must be given by
$X_8$, but assuming it to be true leads some useful clues about the nature of
the spacetime Lorentz-invariant worldvolume action; in order to produce the
mixed anomaly term in $X_8$ some fermion fields must carry {\it
both} d=10 Lorentz and YM indices. This is not necessary for the string because
$X_4$ contains no mixed term.

\bigskip
\centerline{\bf Acknowledgements}
PKT thanks Michael Duff and Andy Strominger for discussions that motivated much
of the work reported here. J.M.I. thanks the Spanish Ministry of Education for
financial support.

\vfill\eject
\centerline{\bf References}
\medskip

\item {[1]}
C.G. Callan and J.A. Harvey, Nucl. Phys. {\bf B250} (1985) 427.

\item {[2]}
S.G. Naculich, Nucl. Phys. {\bf B296} (1988) 837.

\item {[3]}
P.K. Townsend, {\it Effective description of axion defects}, Phys. Lett. {\bf
B} {\it in press}.

\item {[4]}
P. Sikivie, Phys. Lett. {\bf 137B} (1984) 353.

\item {[5]}
K. Lee, Phys. Rev. {\bf D35} (1987) 3286.

\item {[6]}
F. Wilczek, Phys. Rev. Lett. {\bf 58} (1987) 1799.

\item {[7]}
C.M. Hull and E. Witten, Phys. Lett. {\bf 160B} (1985) 398.

\item {[8]}
C. Callan, J. Harvey and A. Strominger, Nucl. Phys. {\bf B359} (1991)
611.

\item {[9]}
M.J. Duff and X. Lu, Nucl. Phys. {\bf B354} (1991) 141.

\item {[10]}
H. Nicolai and P.K. Townsend, Phys. Lett. {\bf 98B} (1981) 257.

\item {[11]}
S. Deser, R. Jackiw and S. Templeton, Phys. Rev. Lett. {\bf 48} (1982) 975;
Ann.
Phys. (N.Y.) {\bf 140} (1982) 372.

\item {[12]}
E. Witten, Nucl. Phys. {\bf B 223} (1983) 422.

\item {[13]}
A. Manohar and G. Moore, Nucl. Phys. {\bf B243} (1984) 55;
\"O. Kaymakcalan, S. Rajeev and J. Schechter, Phys. Rev. {\bf D 30}
594 (1984); J. Ma{\~ n}es, Nucl. Phys. {\bf B250} (1985) 369.

\item {[14]}
N.K. Nielsen, Nucl. Phys. {\bf B167} (1980) 249;
M.J. Duff, B.E.W. Nilsson and C.N. Pope, Phys. Lett. {\bf 163B} (1985)
343; M.J. Duff, B.E.W. Nilsson, C.N. Pope and N.P. Warner, Phys. Lett.
{\bf 171B} (1986) 170; R. Nepomechie, Phys. Lett. {\bf 171B} (1986) 195;
Phys. Rev. {\bf D33} (1986) 3670;
R. Kallosh, Phys. Scripta {\bf T15} (1987) 118;
Phys. Lett. {\bf 176B} (1986) 50.

\item {[15]}
E. Bergshoeff, F. Delduc and E. Sokatchev, Phys. Lett. {\bf 262B}
(1991) 444; P.S. Howe, Phys. Lett. {\bf 273B} (1991) 90.

\item {[16]}
E. Bergshoeff, R. Percacci, E. Sezgin and K.S. Stelle and P.K. Townsend,
Nucl. Phys. {\bf B398} (1993) 343.

\item {[17]}
A. Strominger, Nucl. Phys. {\bf B343} (1990) 167.

\item {[18]}
P.K. Townsend, in {\it Recent Problems in Mathematical Physics}, eds. L.A.
Ibort and M.A. Rodr{\' {\i}}guez, (Kl{\" u}wer 1993).

\item {[19]}
G. Horowitz and A. Strominger, Nucl. Phys. {\bf B360} (1991) 197.

\item {[20]}
M.J. Duff and X. Lu, Phys. Lett. {\bf 273B} (1991) 409.

\item {[21]}
G.W. Gibbons and P.K. Townsend {\it Vacuum interpolation in supergravity via
p-branes}, preprint DAMTP/R-93/19

\item {[22]}
A. Dabholkar, G.W. Gibbons, J.A. Harvey and F. Ruiz-Ruiz, Nucl. Phys. {\bf
B340} (1990) 33.

\item {[23]}
J.A. Dixon, M.J. Duff and E. Sezgin, Phys. Lett. {\bf 279B} (1992)
265.

\item {[24]}
J.A. Dixon, M.J. Duff and J.C. Plefka, Phys. Rev. Lett. {\bf 69} (1992) 3009.

\item {[25]}
L. {\' A}lvarez-Gaum{\' e} and E. Witten, Nucl. Phys. {\bf B234} (1984) 269.

\end